\input{aipcheck}

\documentclass[
    ,final            
  ]
  {aipproc}

\layoutstyle{8x11single}

\usepackage{color}

\usepackage{amssymb}
\usepackage{amsmath}
\begin{document}

\quad{PI/UAN-2013-558FT}

\title{THE DIFFERENT VARIETIES OF THE SUYAMA-YAMAGUCHI CONSISTENCY RELATION}


\author{Yeinzon Rodr\'{\i}guez}{
  address={Centro de Investigaciones en Ciencias B\'asicas y Aplicadas, Universidad Antonio Nari\~no, \\
  Cra 3 Este \# 47A-15, Bogot\'a D.C. 110231, Colombia.}
,altaddress={Escuela de F\'isica, Universidad Industrial de Santander, \\
  Ciudad Universitaria, Bucaramanga 680002, Colombia.}
}



\begin{abstract}
We present the different consistency relations that can be seen as variations of the well known Suyama-Yamaguchi (SY) consistency relation  $\tau_{\rm NL} \geqslant \left(\frac{6}{5} f_{\rm NL} \right)^2$, the latter involving the levels of non-gaussianity $f_{\rm NL}$ and $\tau_{\rm NL}$ in the primordial curvature perturbation $\zeta$, they being scale-invariant.  We explicitly state under which conditions the SY consistency relation has been claimed to hold in its different varieties (implicitly) presented in the literature since its inception back in 2008; as a result, we show for the first time that the variety $\tau_{\rm NL} ({\bf k}_1, {\bf k}_1) \geqslant \left(\frac{6}{5} f_{\rm NL} ({\bf k}_1) \right)^2$, which we call ``the fifth variety'', is always satisfied even when there is strong scale-dependence and high levels of statistical anisotropy as long as statistical homogeneity holds: thus, an observed violation of this specific variety would prevent the comparison between theory and observation, shaking this way the foundations of cosmology as a science.

\end{abstract}

\maketitle


\section{Introduction}
Modern cosmology not only studies the background dynamics of any inflationary model of the Universe, but also the perturbation dynamics via the connected $n$-point correlators $\langle \zeta({\bf k}_1) \zeta({\bf k}_2) ... \zeta({\bf k}_n) \rangle_c$ of the primordial curvature perturbation $\zeta$ \cite{lythbook}.  During some time, the two-point correlator of $\zeta$ was enough for the purpose of comparing theoretical predictions with observation, but, after satellite missions reached amazing accuracy levels \cite{wmap,planck}, it has been necessary to work out the connected three- and four-point correlators of $\zeta$.  Non-vanishing connected three- or four-point correlators of $\zeta$ imply that the probability distribution function of the primordial curvature perturbation is non-gaussian \cite{ValenzuelaToledo:2011fj}, that being the reason why the functions $f_{\rm NL} ({\bf k}_1, {\bf k}_2, {\bf k}_3)$, and $\tau_{\rm NL} ({\bf k}_1, {\bf k}_2, {\bf k}_3, {\bf k}_4)$ and $g_{\rm NL} ({\bf k}_1, {\bf k}_2, {\bf k}_3, {\bf k}_4)$, that parameterize the connected three- and four-point correlators of $\zeta$ respectively are called the levels of non-gaussianity.  A few years ago, Suyama and Yamaguchi \cite{Suyama:2007bg} showed that $f_{\rm NL}$ and $\tau_{\rm NL}$ satisfy the consistency relation $\tau_{\rm NL} \geqslant \left(\frac{6}{5} f_{\rm NL} \right)^2$;  in this paper, which is based on Ref. \cite{Rodriguez:2013cj}, we state explicitly the conditions under which this (first) consistency relation is valid and formulate other five ways, or varieties, to express the SY consistency relation, the conditions under which they are valid, and the respective cosmological implications.

\section{The first variety of the SY consistency relation}
In an elegant work by Smith, LoVerde, and Zaldarriaga \cite{cr2}, it was shown that the first variety of the SY consistency relation:
\begin{equation}
\boxed{\tau_{\rm NL} \geqslant \left( \frac{6}{5} f_{\rm NL} \right)^2} \,,  \label{sye}
\end{equation}
is valid
as long as the following conditions are satisfied:
\begin{itemize}
\item {\it Condition 1}: The calculation of $f_{\rm NL}$ and $\tau_{\rm NL}$ is performed non-perturbatively.
\item {\it Condition 2}: The inflationary dynamics is arbitrary.
\item {\it Condition 2a (actually, a consequence of condition 2)}: The fields involved (if any) can be non-gaussian.
\item {\it Condition 3}: Statistical homogeneity of $\zeta$ is preserved.
\item {\it Condition 4}: Statistical isotropy of $\zeta$ is preserved.
\item {\it Condition 5}: $f_{\rm NL} ({\bf k}_1, {\bf k}_2, {\bf k}_3)$ is evaluated in the squeezed limit (${\bf k}_1 \rightarrow 0$) while $\tau_{\rm NL} ({\bf k}_1, {\bf k}_2, {\bf k}_3, {\bf k}_4)$ is evaluated in the collapsed limit (${\bf k}_1 + {\bf k}_2 \rightarrow 0$).
\item {\it Condition 6}: $f_{\rm NL}$ and $\tau_{\rm NL}$ are scale-invariant.
\end{itemize}
Conditions 2 and 2a may not sound really as conditions in the sense that they should imply some restrictions or hypothesis;  however, they are important because imply that the first variety of the SY consistency relation is generic for all kind of inflationary models as long as the other conditions are satisfied.
Besides, although $f_{\rm NL}$ and $\tau_{\rm NL}$ are not directly comparable in the general case because they are functions of the wavevectors, the latter expression is valid since, under the conditions previously stated, $f_{\rm NL}$ and $\tau_{\rm NL}$ are scale-invariant \cite{Lyth:2005fi}.

The statistical homogeneity of $\zeta$ states that the connected $n$-point correlators in the configuration space are invariant under space translations \cite{ValenzuelaToledo:2011fj};  this implies that, in the momentum space, the connected $n$-point correlators are proportional to a Dirac delta function:
\begin{equation}
\langle \zeta({\bf k}_1) \zeta({\bf k}_2) ... \zeta({\bf k}_n) \rangle_c \propto \delta^3 ({\bf k}_1 + {\bf k}_2 + ... + {\bf k}_n) \,.
\end{equation}
The above implies that the wavevector configuration is such that ${\bf k}_1$, ${\bf k}_2$, ..., and ${\bf k}_n$ form a warped polygon.  The statistical homogeneity is fundamental in cosmology since it allows us to compare theoretical predictions with observations (via the ergodic theorem, see Refs. \cite{ValenzuelaToledo:2011fj,weinbergbook}).  In contrast, the statistical isotropy of $\zeta$ states that the connected $n$-point correlators in the configuration space are invariant under space rotations \cite{ValenzuelaToledo:2011fj};  this implies that, in the momentum space, and once statistical homogeneity has been imposed, each connected $n$-point correlator is proportional to a function $M_\zeta$ called the $n-1$-spectrum which is invariant under rotations in the momentum space:
\begin{equation}
\langle \zeta({\bf k}_1) \zeta({\bf k}_2) ... \zeta({\bf k}_n) \rangle_c = (2\pi)^3 \delta^3 ({\bf k}_1 + {\bf k}_2 + ... + {\bf k}_n)  M_\zeta ({\bf k}_1, {\bf k}_2, ... , {\bf k}_n) \,,
\end{equation}
where
\begin{equation}
M_\zeta (\tilde{\bf k}_1, \tilde{\bf k}_2, ... ,\tilde{\bf k}_n) = M_\zeta ({\bf k}_1, {\bf k}_2, ... , {\bf k}_n) \,,  \label{spectra}
\end{equation}
being $\tilde{\bf k}_i = \mathcal{R} \ {\bf k}_i$ with $\mathcal{R}$ being the rotation operator such that $\tilde{\bf x}_i = \mathcal{R} \ {\bf x}_i$ in the configuration space, i.e. the $n-1$-spectrum is the same no matter the orientation of the warped polygon.  The statistical isotropy is the usual assumption in cosmology since the inflationary dynamics is usually driven by scalar fields only, which, due to its nature, do not exhibit preferred directions.

\section{The second and third varieties of the SY consistency relation}
What happens if Conditions 4 and 6 in the first variety of the SY consistency relation do not hold?  Well, the first variety then reduces to what we call the second variety of the SY consistency relation \cite{cr2}:
\begin{equation}
\boxed{\int_{{\bf k}_1, {\bf k}_3 \in b_s} \frac{d^3 {\bf k}_1 d^3 {\bf k}_3}{(2\pi)^6} \tau_{\rm NL} ({\bf k}_1, {\bf k}_3) \geqslant \left[ \frac{6}{5} \int_{{\bf k}_2 \in b_s} \frac{d^3 {\bf k}_2}{(2\pi)^3}  f_{\rm NL} ({\bf k}_2) \right]^2} \,, \label{slzconsistency}
\end{equation}
where $b_S$ is a 
narrow band of wavevectors which are very
near some ${\bf k}_S$.  This ${\bf k}_S$ is actually arbitrary as well as the width of $b_S$.

By employing an alternative technique, Assassi, Baumann, and Green \cite{abg} obtained a similar result under the same assumptions;  this is what we call the third variety of the SY consistency relation:
\begin{equation}
\boxed{\int \frac{d^3{\bf k}_1 d^3{\bf k}_3}{(2\pi)^6} \tau_{\rm NL} ({\bf k}_1, {\bf k}_3) P_\zeta ({\bf k}_1) P_\zeta ({\bf k}_3)  \geqslant \left[ \frac{6}{5} \int \frac{d^3 {\bf k}_2}{(2\pi)^3}  f_{\rm NL} ({\bf k}_2) P_\zeta ({\bf k}_2) \right]^2} \,,  \label{abgconsistency}
\end{equation}
where $P_\zeta ({\bf k})$ is the power spectrum of the primordial curvature perturbation $\zeta$.

Although one cannot neglect the power of a consistency relation, the second and third varieties presented above are not well suited to confront with observations because of the momentum integrals involved ($f_{\rm NL} ({\bf k}_2)$ and $\tau_{\rm NL} ({\bf k}_1, {\bf k}_3)$ are actually not measured for the whole momentum space);  anyway, we will explicitly state the conditions under which these varieties are satisfied:
\begin{itemize}
\item {\it Condition 1}: The calculation of $f_{\rm NL}$ and $\tau_{\rm NL}$ is performed non-perturbatively.
\item {\it Condition 2}: The inflationary dynamics is arbitrary.
\item {\it Condition 2a (actually, a consequence of condition 2)}: The fields involved (if any) can be non-gaussian.
\item {\it Condition 3}: Statistical homogeneity of $\zeta$ is preserved.
\item {\it Condition 4}: $f_{\rm NL} ({\bf k}_1, {\bf k}_2, {\bf k}_3)$ is evaluated in the squeezed limit (${\bf k}_1 \rightarrow 0$) while $\tau_{\rm NL} ({\bf k}_1, {\bf k}_2, {\bf k}_3, {\bf k}_4)$ is evaluated in the collapsed limit (${\bf k}_1 + {\bf k}_2 \rightarrow 0$).
\end{itemize}

\section{The fourth and fifth varieties of the SY consistency relation}
Looking at the second and third varieties of the SY consistency relation, we can propose a direct generalization of the first variety, Eq. (\ref{sye}), when there is no scale-invariance and whose form is easily inspired from the second and third varieties, Eqs. (\ref{slzconsistency}) and (\ref{abgconsistency}):
\begin{equation}
\boxed{\tau_{\rm NL} ({\bf k}_1, {\bf k}_3) \geqslant \left(\frac{6}{5}\right)^2 f_{\rm NL} ({\bf k}_1) f_{\rm NL} ({\bf k}_3)} \,. \label{mpsy}
\end{equation}
This is what we call the fourth variety of the SY consistency relation.  There are no conditions for this variety since it is just a proposal;  thus, we are sure that this variety is strongly violated for models that involve non-trivial degrees of freedom;  indeed, examples of such a violation are presented in Refs. \cite{Rodriguez:2013cj,kr,Choudhury:2012kw,Shiraishi:2013vja}.

Meanwhile, what we can say is that the fourth variety of the SY consistency relation is satisfied when the arguments of $\tau_{\rm NL}$ and $f_{\rm NL}$  are all the same even if the scale invariance is not guaranteed:
\begin{equation}
\boxed{\tau_{\rm NL} ({\bf k}_1, {\bf k}_1) \geqslant \left(\frac{6}{5} f_{\rm NL} ({\bf k}_1) \right)^2} \,; \label{5vsy}
\end{equation}
this can be proven by observing that, in Eq. (\ref{slzconsistency}), $b_S$ can be made as small as we want, getting rid of the integrals, but making the arguments of $\tau_{\rm NL}$ and $f_{\rm NL}$ all the same. We will call Eq. (\ref{5vsy}) the fifth variety of the SY consistency relation. Since this fifth variety is derived from Eq. (\ref{slzconsistency}), the conditions under which it holds are the Conditions 1 to 4 stated at the end of the previous section.
As seen there, the statistical homogeneity is actually the only non-general condition for the validity of the fifth variety of the SY consistency relation when $\tau_{\rm NL} ({\bf k}_1, {\bf k}_2, {\bf k}_3, {\bf k}_4)$ and $f_{\rm NL} ({\bf k}_1, {\bf k}_2, {\bf k}_3)$ are evaluated in the collapsed and squeezed limits respectively; an observed violation of such a variety would lead to the conclusion that statistical homogeneity does not hold and that, therefore, theory and observation cannot be compared \cite{ValenzuelaToledo:2011fj} leading to a strong shaking of the foundations of cosmology as a science \cite{yrodrig}.

\section{The sixth variety of the SY consistency relation}
We briefly present in this section the latest variety of the SY consistency relation which was formally demonstrated in a quite recent paper by the author of this work and his colleagues \cite{BeltranAlmeida:2011db}:
\begin{equation}
\boxed{\tau^{\rm iso}_{\rm NL} \geqslant \left( \frac{6}{5} f^{\rm iso}_{\rm NL} \right)^2} \,.  \label{syvf}
\end{equation}
In the previous expression, $\tau^{\rm iso}_{\rm NL}$ and $f^{\rm iso}_{\rm NL}$ correspond to the isotropic pieces (i.e., they do not depend at all on the wavevectors) of the levels of non-gaussianity in models where the primordial curvature perturbation is generated by scalar and vector fields. One could think that $\tau^{\rm iso}_{\rm NL}$ and $f^{\rm iso}_{\rm NL}$, being isotropic pieces, have contributions only from the scalar fields, but, as demonstrated in \cite{BeltranAlmeida:2011db}, they receive contributions from the vector fields as well.

We will enumerate the conditions under which the sixth variety of the SY consistency relation is satisfied:
\begin{itemize}
\item {\it Condition 1}: The calculation of $f^{\rm iso}_{\rm NL}$ and $\tau^{\rm iso}_{\rm NL}$ is performed at tree level in the diagrammatic approach of the $\delta N$ formalism \cite{ValenzuelaToledo:2011fj,bksw}.
\item {\it Condition 2}: The expansion is assumed to be isotropic so that the field perturbations (actually the scalar perturbations that multiply the respective polarization vectors (if any) in a polarization mode expansion) are statistically isotropic.
\item {\it Condition 3}: The inflationary dynamics is driven by any number of slowly-rolling vector and scalar fields.
\item {\it Condition 4}: The fields involved are gaussian.
\item {\it Condition 5}: The field perturbations (with the proviso expressed in {\it Condition 2}) are scale-invariant.
\end{itemize}
As discussed previously, $f_{\rm NL}$ and $\tau_{\rm NL}$ are not directly comparable in the general case because they are functions of the wavevectors; however, the expression in Eq. (\ref{syvf}) is valid since, under the conditions just stated above, $f^{\rm iso}_{\rm NL}$ and $\tau^{\rm iso}_{\rm NL}$ are scale-invariant \cite{dklr}.

\section{Conclusions}
When first introduced back in 2008 \cite{Suyama:2007bg}, the SY consistency relation was considered as a useful tool to discriminate among different classes of models for the generation of the primordial curvature perturbation;  however, the quite strong conditions under which this relation was proven led several authors to wonder whether such conditions could be relaxed in the search of a more accurate and clean way to employ this consistency relation as a discriminator tool.  Since that time, six different varieties of the SY consistency relation have been implicitly introduced in the literature, one of them indeed introduced in the present paper.  In this work, we have collected the six varieties and showed them explicitly as well as the conditions under which they are valid.  The first and the sixth varieties, Eqs. (\ref{sye}) and (\ref{syvf}), are quite similar in the sense that the non-gaussianity parameters $f_{\rm NL}$ and $\tau_{\rm NL}$ are scale-invariant;  they are different in the sense that the first variety involves the whole $f_{\rm NL}$ and $\tau_{\rm NL}$ while the sixth variety involves only their isotropic pieces.  The second and third varieties of the SY consistency relation, Eqs. (\ref{slzconsistency}) and (\ref{abgconsistency}), are very similar since they involve an integration over the momentum space but are different because the integration in the second variety is just in a narrow band of wavevectors whereas in the third variety the integration is over the entire momentum space while introducing the power spectrum of the primordial curvature perturbation as part of the integrand;  anyway, because of the integrations, it looks quite difficult to employ these varieties of the SY consistency relation in order to discriminate among different classes of models.  What is interesting however of these varieties is that they are valid even if there is neither statistical isotropy nor scale-invariance:  that is why the fourth variety in Eq. (\ref{mpsy}) is so interesting, being just the best-defined generalization of the first variety and whose form is easily inspired from the second and third varieties;  it is for sure valid for models involving only slowly-rolling Lorentz-invariant scalar fields, as long as the proof by Kehagias and Riotto \cite{kr} turns out to be valid at all orders in perturbation theory, but for sure fails in some model that involves non-trivial degrees of freedom;  indeed, Refs. \cite{Rodriguez:2013cj,kr,Choudhury:2012kw,Shiraishi:2013vja}
have shown the first counterexamples. The latter results sound ubiquitous for any variety of the SY consistency relation, that does not involve integrations over momentum space, when there is no scale-invariance; nonetheless, we have been able to introduce the fifth variety of the SY consistency relation in Eq. (\ref{5vsy}) whose validity is only restricted to the existence of statistical homogeneity and when $f_{\rm NL} ({\bf k}_1, {\bf k}_2, {\bf k}_3)$ and $\tau_{\rm NL} ({\bf k}_1, {\bf k}_2, {\bf k}_3, {\bf k}_4)$ are calculated in the squeezed and collapsed limits respectively;  perhaps the strongest conclusion we can extract from all this analysis is a simple question:  what will we do if the fifth variety of the SY consistency relation is observationally violated?  It would imply statistical inhomogeneity, which in turn implies that, although we dispose of very good data and very good theory, we cannot compare them to each other \cite{ValenzuelaToledo:2011fj,weinbergbook}:  this situation would be a real shake in the foundations of cosmology as a science \cite{yrodrig}.


\begin{theacknowledgments}
The author would like to thank the organizers of the IX Mexican School of the DGFM-SMF for his hospitality and attention. He also acknowledges support for mobility from VCTI (UAN). The author is supported by Fundaci\'on para la Promoci\'on de la Investigaci\'on y la Tecnolog\'{\i}a del Banco de la Rep\'ublica (COLOMBIA) grant number 3025 CT-2012-02, VCTI (UAN) grant number 2011254, and DIEF de Ciencias (UIS) grant number 5177. 
\end{theacknowledgments}



\bibliographystyle{aipproc}   





\end{document}